\documentstyle[proceedings]{crckapb}
\begin{opening}
\title{PULSATING STELLAR ATMOSPHERES}
\author{Dimitar D. Sasselov}
\institute{Harvard-Smithsonian Center for Astrophysics, Cambridge,\\
 MA 02138, USA}
\end{opening}
\begin{document}

\begin{abstract}
We review the basic concepts, present state of theoretical models, and the
future prospects for theory and observations of pulsating stellar atmospheres.
Our emphasis is on radially pulsating cool
stars, which dynamic atmospheres provide a general example for the
differences with standard static model atmospheres.
\end{abstract}

\section{Introduction}

Most stars in the upper part of the Hertzsprung-Russell diagram
pulsate. This pulsation involves large-scale cyclic motion of their envelopes,
very often in a simple radial mode. Most often the stellar atmosphere has
little effect on the driving and acoustics of the pulsation, which is
in essence an envelope phenomenon. However, apart from providing the
upper boundary condition, the atmosphere is what we observe. Except for the
period of pulsation, all other stellar parameters we derive are affected
by the ever changing state of the atmosphere. 

There is a very large range in pulsation amplitudes 
$-$ from the pervasive, but very calm, solar-type
oscillations to radial pulsation of up to a fifth of the stellar radius
in RR Lyrae stars. A large number of giants and supergiants
exhibit cyclic variations of low amplitude which are most likely due
to pulsation (Rao et al. 1993). Even stars which are often considered to
be non-variable standards, like $\gamma$ Cyg and $\alpha$ Per, show such
variability when subjected to scrutiny (Butler 1997). 
There are specific classes of pulsating stars of great interest
to distance measurement and stellar evolution, e.g. Cepheids, RR Lyrae,
Miras, and post-AGB stars, where understanding their pulsating atmospheres is
often crucial to the applications.

\medskip

\noindent {------------} \\
\noindent ${\em in~Proc.~IAU~Symp.~189,~eds.~T.~Bedding~et~al.,~Kluwer,~p.253,~1997}$.

\section{Concepts}

In the talks and discussions of this session, the physics of
stellar atmospheres was discussed explicitly or implicitly in the framework
of some basic approximations. Hydrostatic equilibrium [HSE] is one of them, 
and a very well justified one at that. Here the basic physics of pulsating
atmospheres can be illustrated with the help of juxtapositions and
differences.

\subsection{HSE vs. non-HSE}

The atmosphere of a pulsating star is not in HSE by definition. However, there
is no general and sharply defined state at which HSE breaks down as a viable
approximation. One obvious reason for this is that the amplitude of a global
radial or nonradial oscillation is not the only parameter involved in reshaping
the atmosphere $-$ the period (i.e. velocity gradients) and the state of the
atmosphere are equally important. The latter will determine the acoustic and
hydrodynamic properties of the atmosphere and its response to the oscillation
at its inner boundary.

Pulsation (and departure from HSE) leads to ${\bf extension}$ of the
stellar atmosphere. This change in atmospheric structure is obvious, though
not trivial (e.g. Bowen 1988; Cuntz 1989).  Extension alone 
affects spectral line formation, invalidates the plane-parallel approximation,
and facilitates dust formation and mass loss, as we describe below.

\begin{figure}
\includegraphics{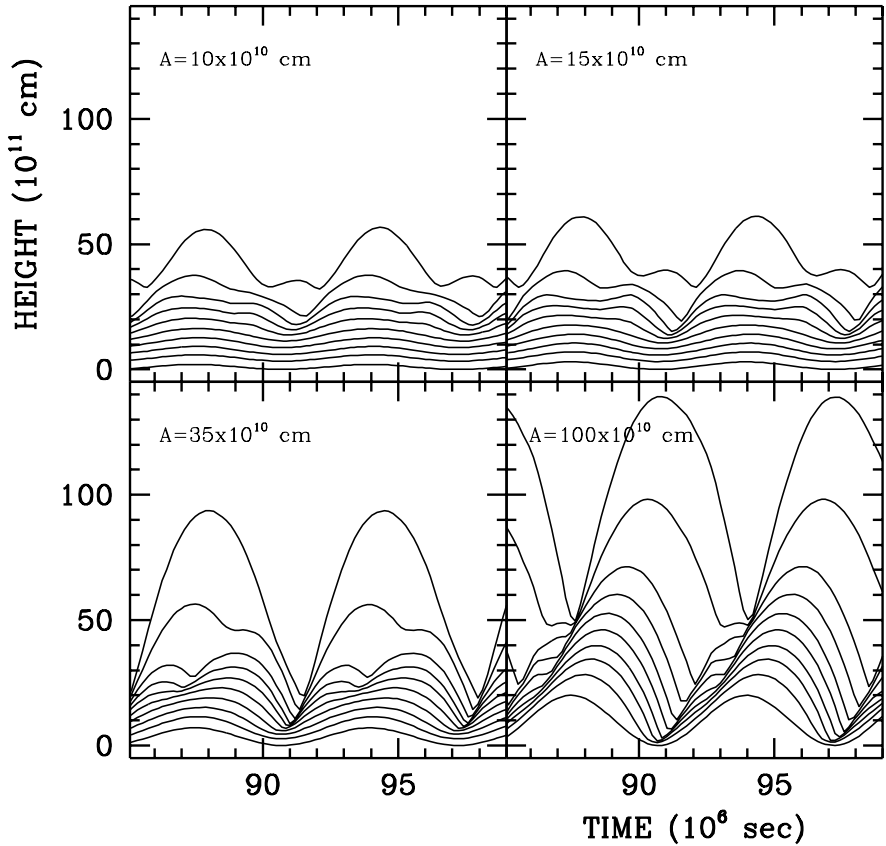}
\vspace*{4.7in}
\caption{The effect of increasing the amplitude of pulsation in an atmosphere 
initially in HSE.
The motion of selected mass zones in the pulsating atmosphere of
a bright giant for four different amplitudes of radial pulsation. The atmosphere
extends by a factor of 2 in the illustrated 
range of piston amplitudes; however, the relation is not linear.}
\end{figure}

The development of extension in an atmosphere undergoing radial pulsation is
shown in Figure 1. The model is for a bright giant or supergiant of $T_{\rm eff}
\approx 5500K$. The pulsation period (75 days) and amplitudes (0.5 to 2 
$km~s^{-1}$) match those observed in stars like $\gamma$ Cyg and $\alpha$ Per.
The model is a 1-D radiation hydrodynamics calculation with H, He, and CaII in
non-LTE (using code 
HERMES by Sasselov \& Raga 1992). Two effects are illustrated:
(1) the general extension of the atmosphere, and (2) the strong shock waves
forming in the upper atmosphere when velocity gradients are too weak to induce
resonant line cooling in the higher-density photosphere.
Shock dynamics is discussed in the next section.

Atmospheric extension affects the formation of ${\bf photospheric~lines}$.
With the accompanying changes in temperature and pressure
structure, extension broadens and complicates the line forming region
(Figure 2). This does not always cause departures from local thermodynamic
equilibrium [LTE] $-$ the transition in Figure 2 (taken from the Cepheid model
of Sabbey ${\it et al.}$ 1995) is formed very close to LTE 
in both the HSE and non-HSE versons of the same model.
However, the resulting spectral line becomes more sensitive to depth-dependent
perturbations.

\begin{figure}
\includegraphics{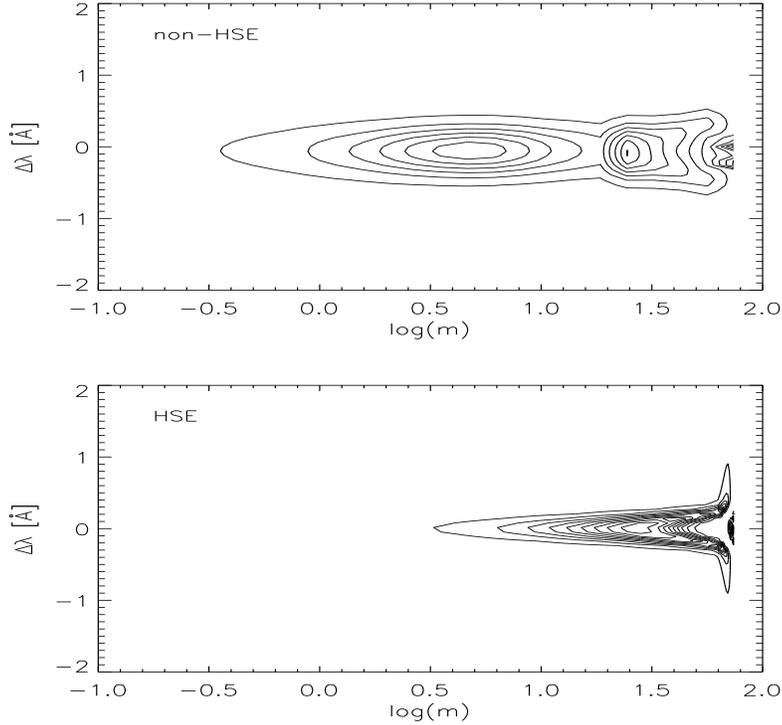}
\vspace*{4.0in}
\caption{The effect of departure from HSE on the formation of the 
Mg~II $\lambda$10952 line (${\em upper~panel}$). The line forming regions
are represented by the contribution function 
(erg cm$^{-2}$ s$^{-1}$ Hz$^{-1}$ sr$^{-1}$) of the Mg~II transition in
wavelength and column mass, $m$ (g cm$^{-2}$).}
\end{figure}

The ${\bf plane}$-${\bf parallel}$ approximation 
fails when atmospheric
extension exceeds $\approx$5\% of the stellar radius (Schmid-Burgk \&
Scholz 1975). The equation of radiative transfer has to be solved in
spherical geometry. The radiation field is diluted throughout the atmosphere,
the temperature gradient becomes steeper, and the emergent flux distribution
changes (it becomes flatter from UV to IR). Hence colour indices between
blue, red, and near-IR bandpasses are underestimated (Fieldus \& Lester 1990).
The line spectrum changes too: a steeper source function tends to enhance
most weak absorption lines.

With the drop in both temperature and effective gravity
as a function of height, atmospheric extension creates better conditions
for the formation of ${\bf dust}$ (Gauger, Sedlmayr, \& Gail 1990) 
and for ${\bf mass~loss}$ (Hoefner ${\it et al.}$ 1996).

\subsection{Radiative equilibrium vs. Shock Waves}

Radiative equilibrium [RE] is unlikely in a dynamic and extended stellar
atmosphere. Neither is convective equlibrium (see Kurucz, \S~2, this volume),
although both radiative and convective energy transport remain 
the most important.
Dynamics and the development of shock waves introduce complexities in the
atmosphere which affect both spectral lines and continuum emission.

The mechanical flux deposited into the atmosphere by the pulsation (and
the dynamics/shocks induced by it) will affect its thermodynamical state.
In cool star atmospheres this heating could go into ionization of H and
He, and be lost radiatively in resonance lines, etc. It is thus important to
treat the solution in non-LTE (Carlsson \& Stein 1992).
Emergent continuum radiation will be less affected than spectral lines
due to the non-locality of the source function and the larger range of depths
involved.

The large-scale dynamics resulting from the pulsation is strongly coupled
and affected by the ${\bf small}$-${\bf scale~nonthermal~velocity~fields}$
present in the atmosphere.
The origin and nature of the small-scale velocity fields are probably
unrelated to the large-scale radial pulsation motion. Instead, the solar-stellar
connection points towards convective motions and high-frequency waves as
being responsible (see Cottrell, this volume).
Traditionally, photon pathlengths have been used to define the difference
between large-scale and small-scale velocity fields in a stellar atmosphere.
These are called macroturbulent and microturbulent velocity, respectively
$-$ $ad~hoc$ parameters not to be identified with hydrodynamic turbulence,
but with extremes of spectral line broadening parametrization (Mihalas 1978).
In a standard static atmosphere model the turbulent velocity is an empirical
constant parameter.

The need for a $depth$-$dependent$
microturbulent velocity (also, time-dependent)
arises clearly in pulsating atmospheres, when hydrodynamics and radiative
transfer are
coupled (Sasselov \& Lester 1994). This is important for synthesizing
realistic line profiles, but also because more microturbulent velocity
broadening in cool stars reduces atmospheric opacity by reducing the gas
pressure contribution. As regards macroturbulent velocity, there has been
no need for introducing it in HERMES models of Cepheids, because
such broadening is consistently
computed in terms of the large-scale motions induced by the pulsation.
The same set of hydrodynamic equations should also be adequate to compute the
small-scale velocity fields if not for: (1) limited grid resolution (compared
to photon mean free paths), and (2) lack of a mechanism for energy transport
between large and small scales (e.g. turbulent cascade) in a 1-D model.
Undertanding the mechanism for energy transport could solve the problem
even for 1-D codes. For example, assuming that small-scale velocity fields
are well represented by a turbulent spectrum one could model the 
depth-dependent microturbulent velocity with a simple self-consistent model
of turbulence in the spirit of Canuto, Goldman, \& Mazzitelli (1996). Thus
one would treat both the subgrid cascade, as well as the nonlinear interaction
of the pulsation shock waves with the small-scale field.

\section{Current State of Theory}

\subsection{Models}
Computing pulsating stellar atmospheres across the HR diagram poses
challenges of different nature in different types of stars. In some
the requirements towards the hydrodynamics are higher, in others $-$
the radiative transfer is more important. Therefore, currently there
are three loosely defined types of models for pulsating atmospheres.
(We limit references to most recent models only.)

${\bf Envelope~models}$ have an inner boundary at a few 10$^6$K  and calculate
the driving and acoustics of the pulsation with a first-order hydro scheme.
The pulsating atmosphere is calculated (in LTE or non-LTE) after the run
of the envelope model from selected snapshots. Such models are useful for
both hotter pulsators (like Cepheids) and cooler ones (like Miras),
and especially for the latter (e.g. Bessell, Scholz, \& Wood 1996;
Luttermoser \& Bowen 1990). 
For Cepheids and similar stars $-$ models calculations
by Fokin (1991); Albrow \& Cottrell (1996). All these models are in 1-D.

${\bf Gray~atmosphere~models}$ make use of a piston at their inner boundary
(at $\approx$10$^4$K) to introduce the pulsation motion. At their outer
boundary the radiative transfer is calculated assuming constant opacity
(gray approximation). Despite the strong limitations of the gray
approximation, these codes are useful for studies of the bulk dynamics of
very cool pulsators in 1-D, like Miras (Bowen 1988;
Feuchtinger, Dorfi, \& Hoefner 1993 - who use the Eddington approximation).

${\bf Multi}$-${\bf level~non}$-${\bf LTE~hydro~models}$ also make 
use of a piston at
their inner boundary (at $\approx$10$^4$K). They use second order
accurate (no artificial viscosity) hydro schemes with multi-level
non-LTE radiative transfer for cooling and heating terms in the energy
equation. Standard assumptions in such models are: (1) that non-LTE
does not affect the hydro solution for the density and velocity;
and (2) time dependence for bound-free transitions (assuming that
the bound-bound rates are much larger). Models based on the method
of characteristics were developed by Cuntz (1989).
Sasselov \& Raga (1992) use the Godunov method and multi-level
atoms of H, He, Ca, and Mg. These codes are also in 1-D.

\subsection{Challenges \& Problems}
All problems of standard stellar model atmospheres (see Gustafsson,
Kurucz, this volume) apply also to pulsating stellar atmospheres.
In addition, there are some important bits of physics which are still
left out of pulsating model atmospheres.

${\em Line~blanketing}$ is not yet incorporated fully (or at all) and
all current experience from static model atmospheres points to its importance.
This means that we are currently unable to judge the systematic errors in
the use of colors and color calibrations to pulsating stars. There is no
unsurmountable obstacle in building a line-blanketed 1-D raditive
hydrodynamics model for hotter variables (like Cepheids). This should be
done, given the recent interest in metallicity dependence of Cepheid
distances, which relies strongly in understanding colors and color changes.

${\em Important~coolants}$ are still missing from the coupled non-LTE
hydrodynamics models, in particular - FeI and FeII. While Fe species
are included very roughly in the background opacities, short of a
full line blanketing calculation, non-LTE coupling of Fe may be as
important as that of He, Ca, and Mg, which are currently used.
Including Fe in non-LTE models is also within reach, given the successes
of novae and supernovae models.

${\em Shock~precursors}$ and the treatment of strong shocks in general
require significant improvement. This is a difficult problem, as it
touches upon the isssues discussed above, as well as the physically
consistent treatment of shock-turbulence interaction. On one hand,
the recurrence of shocks in a pulsating atmosphere could lead to enhancement
of the small-scale nonthermal velocity fields $-$ a typical shock strength
in a Cepheid model would cause up to a factor of 2 increase in the turbulent
kinetic energy. In a fluid dynamics sense the problem has been studied by
e.g. Rotman (1991); some observational evidence comes from e.g. Breitfellner
\& Gillet (1994). Another issue is the effect these small-scale velocity fields
have on the front of the shock wave $-$ under most stellar atmosphere conditions
the shock front will be corrugated, thus affecting significantly the energetics
and the precursor solution. In 1-D models this can be handled approximately
by using simple relations for the development of Richtmeyer-Meshkov [RM]
instabilities in the linear regime. One can treat RM instabilities in the
spirit of the Rayleigh-Taylor intability with the shock as an instantaneous
acceleration (Mikaelian 1991). However, it is highly questionable whether
this approach is justified in a 1-D model, where also the small-scale
velocity fields have to be calculated with the same underlying approximation.
Clearly a 2-D or 3-D model is the only consistent solution.

\section{The Future}

Many of the advances in understanding pulsating stellar atmospheres will
follow or parallel advances in standard atmospheres and the Sun. Among them
are the physics of small-scale nonthermal velocity fields and the use of
2-D and 3-D models. Currently there are no outstanding theoretical problems
in coupling radiative transfer and hydrodynamics in 3-D. There has been
considerable progress
in radiative transfer techniques $-$ to mention a few: the ALI
and the MALI methods (Rybicki \& Hummer 1991;
Auer $etal.$ 1994). Fast or refined hydro schemes $-$ e.g. PPM and
Godunov methods (Zachary $etal.$ 1994) are also available.

The ultimate model for a pulsating stellar atmosphere $-$ the $unified$
$pulsation$ $model$, has been a goal for decades. Its completion (even in a 1-D)
will most likely take us beyond the year 2000. The task is complex $-$
a model which couples consistently a realistic pulsating envelope and a
realistic non-LTE line-blanketed atmosphere requires a solution to at least
three outstanding problems. The first problem is time-dependent convection.
The second problem is the smooth transition between table opacities (in the
envelope) and multi-level radiative transfer (in the atmosphere). This
transition occurs in a very sensitive location in the envelope-atmosphere
interface (with effects on the acoustic cavity and the convection zone).
In order to tackle this problem we need a ``smarter" adaptive grid method
which can manage efficiently the often opposing demands of the hydrodynamics
and the non-LTE radiative transfer. Building such an adaptive grid is the
third outstanding problem $-$ it is not simply an issue of enough resolution
(which is sorely needed!),
but rather of being able to handle the approximate physics in the
envelope-atmosphere transition region, and keep the two together.
This will be, in the words of Sir Walter Scott, ``${\em The~silver~link}$,
${\em the~silken~tie}$, ${\em Which~heart~to~heart}$, ${\em and~mind~to~mind}$,
${\em In~body~and~in~soul~can~bind}$".

\section{References}

Auer, L., Bendicho, F., \& Bueno, T. 1994, A\&A, 292, 599\\
Albrow, M., \& Cottrell, P. 1996, MNRAS, 280, 917\\
Bessell, M., Scholz, M., \& Wood, P. 1996, A\&A, 307, 481\\
Bowen, G. W. 1988, ApJ, 329, 299\\
Breitfellner, M., \& Gillet, D. 1993, A\&A, 277, 553\\
Butler, P. 1997, preprint\\
Canuto, V. M., Goldman, I., \& Mazzitelli, I. 1996, ApJ, 473, 550\\
Carlsson, M., \& Stein, R. 1992, ApJ, 397, L59\\
Cuntz, M. 1989, PASP, 101, 560\\
Feuchtinger, M., Dorfi, E., \& Hoefner, S. 1993, A\&A, 273, 513\\
Fieldus, M., \& Lester, J. 1990, in ${\it 6th~Cambridge~Workshop}$, ASP 9, 79\\
Fokin, A. B. 1991, MNRAS, 250, 258\\
Gauger, A., Sedlmayr, E., \& Gail, H.P. 1990, A\&A, 235, 345\\
Hoefner, S., ${\it et~al.}$ 1996, A\&A, 314, 204\\
Luttermoser, D., \& Bowen, G. 1990, in ${\it 6th~Cambridge~Worksh.}$,ASP 9,491\\
Mikaelian, K. 1991, Phys. Fluids A, 3, 2625\\
Rybicki, G., \& Hummer, D. 1991, A\&A, 245, 171\\
Rao, L., ${\it et~al.}$ 1993, in ${\it Luminous~High-latitude~Stars}$, ASP 45,
 300\\
Rotman, D. 1991, Phys. Fluids A, 3, 1792\\
Sabbey, C., Sasselov, D., Fieldus, M., Lester, J., Venn, K., \& Butler, P.\\
 \indent{1995, ApJ, 446, 250}\\ 
Sasselov, D., \& Raga, A. 1992, in ${\it 7th~Cambridge~Workshop}$, ASP 26, 549\\
Sasselov, D., \& Lester, J. B. 1994, ApJ, 423, 795\\
Schmid-Burgk, J., \& Scholz, M. 1975, A\&A, 41, 41\\
Zachary, A., Malagoli, A., \& Colella, P. 1994, SIAM J. Sci. Comp., 15, 263.

\medskip
\small
\tt\raggedright
\parindent=0pt
\medskip
{\bf DISCUSSION}
 
\medskip
 
JOHANNES ANDERSEN: Your results seem to demolish most of the assumptions
underlying the simple-minded applications of the Baade-Wesselink method, in
particular the determination of radial-velocity curves from spectra
averaging lots of lines of different origins.  Is there any hope of getting
useful results from these existing observations, or will they all have to
be thrown out and replaced by new data from selected sets of spectral
lines?
 
\medskip
 
DIMITAR SASSELOV: The Baade-Wesselink method is attractive because it has
the potential to provide accurate one-step distances to nearby galaxies.
The simple-minded application of the method suffers from systematic errors
which puts its reliability in question and makes it less competitive with
other methods.  One way to improve the quality of BW solutions is with the
use high-resolution spectra.  Unfortunately, the necessary information on
line profiles and strengths cannot be recovered from existing observations.
 
\medskip
 
MICHAEL SCHOLZ: Perhaps, the situation is different in Cepheids, but in
case of the Miras the structure of the deep atmospheric layers (= upper
interior `envelope') changes little as you replace the simple grey
atmosphere by a more sophisticated non-grey atmosphere.
 
\medskip
 
DIMITAR SASSELOV: Yes, the atmospheres of Miras are so extended that
regions in them may be completely decoupled and behave locally with the
corresponding molecular opacitites.
 
\medskip
 
BENGT GUSTAFSSON: Concerning the effects of blanketing - are they, in
addition to being significant for colour calculations, etc., also
significant for the structure and dynamics of the upper layers?
 
\medskip
 
DIMITAR SASSELOV: Yes, I suspect they may be significant.  We will not know
how significant until a consistent calculation is made.

\end{document}